\documentstyle[editedvolume,psfig,raulbib]{crckapb}
\begin{opening}
\title{Extracting cosmological information from galaxy spectra and
  observations of high-redshift objects}
\author{Raul Jimenez}
\institute{Department of Physics and Astronomy, Rutgers University, \\ 
136 Frelinghuysen Road, Piscataway, NJ 08854--8019 USA \\ 
raulj@physics.rutgers.edu}
\end{opening}

\runningtitle{Extracting cosmological information}

\begin{document}

\newcommand{\be}{\begin{equation}}
\newcommand{\ee}{\end{equation}}
\newcommand{\ba}{\begin{eqnarray}}
\newcommand{\ea}{\end{eqnarray}}
\newcommand{\mat}{{\bf}}
\newcommand{\nn}{\nonumber\\}
\newcommand{\bA}{{\bf A}}
\newcommand{\bE}{{\bf E}}
\newcommand{\bB}{{\bf B}}
\newcommand{\bC}{{\bf C}}
\newcommand{\bF}{{\bf F}}
\newcommand{\bS}{{\bf S}}
\newcommand{\bfa}{{\bf a}}
\newcommand{\bb}{{\bf b}}
\newcommand{\bg}{{\bf g}}
\newcommand{\bj}{{\bf j}}
\newcommand{\bk}{{\bf k}}
\newcommand{\bn}{{\bf n}}
\newcommand{\bp}{{\bf p}}
\newcommand{\br}{{\bf r}}
\newcommand{\bu}{{\bf u}}
\newcommand{\bv}{{\bf v}}
\newcommand{\bx}{{\bf x}}
\newcommand{\by}{{\bf y}}
\newcommand{\bbmu}{\mbox{\boldmath $\mu$}}
\newcommand{\dd}{{\partial}}
\newcommand{\ddt}{{\partial\over \partial t}}
\def\gs{\mathrel{\raise1.16pt\hbox{$>$}\kern-7.0pt 
\lower3.06pt\hbox{{$\scriptstyle \sim$}}}}         
\def\ls{\mathrel{\raise1.16pt\hbox{$<$}\kern-7.0pt 
\lower3.06pt\hbox{{$\scriptstyle \sim$}}}}         

\begin{abstract}
  I review the statistical techniques needed to extract information about
  physical parameters of galaxies from their observed spectra. This is
  important given the sheer size of the next generation of large galaxy
  redshift surveys. Going to the opposite extreme I review what we can learn
  about the nature of the primordial density field from observations of
  high--redshift objects.
\end{abstract}

\section{Extracting cosmological information from galaxy spectra}

Most of the information about the physical properties of galaxies comes from
their electromagnetic spectrum. It is therefore of paramount importance to be
able to extract as much physical information as possible from it. In
principle, it is straightforward to determine physical parameters from an
individual galaxy spectrum. The method consists in building synthetic stellar
population models which cover a large enough range in the parameter space and
then use a merit function (typically a $\chi^2$) to evaluate which suite of
parameters better fits the observed spectrum. There are two obvious
limitations of the above method: first, the number of parameters that govern
the spectrum of a galaxy may be very large and thus difficult to explore
fully. Secondly, in the case of ongoing large redshifts surveys which will
provide us with about a million galaxy spectra, it will be computationally
very expensive  (and possibly intractable for redshift surveys like the 2dF
and SDSS) to apply a plain $\chi^2$ to each individual spectrum which
itself may contain of the order of $10^3$ data points.

The non-obvious route to tackle the problem is to compress the original data
set in order to weight more those pixel in the spectrum that carry most
information about a given parameter. It is worth reminding that non--optimal
data compression is commonly applied to galaxy spectra: photometric filters.
Not surprisingly, this empirical data compression is not optimal since it has
not been devised to be lossless, i.e. contain all the information for a given
parameter. For example, the photometric $B$ filter alone is not optimal to
recover the age of a galaxy. On the other hand, more sophisticated and
non-empirical methods have been proposed for extracting information from
galaxy spectra, some of them as old as the Johnson's filter system itself.
Many of these are based on Principal Component Analysis or wavelet
decomposition
\cite{Murtagh87,Francis92,Connolly95,Folkes96,Galaz98,Bromley98,Glazebrook98,Singh98,Connolly99,Ronen99,Folkes99}.
PCA projects galaxy spectra onto a small number of orthogonal components. The
weighting of each component corresponds to it's relative importance in the
spectra. However while these components appear to correlate well with physical
properties of galaxies, their interpretation is difficult since they do not
have known, specific physical properties; they can be amalgams of different
properties. To interpret these components, we have to return to model spectra
and compare them with the components \cite{Ronen99}. This is a disadvantage of
PCA since one important goal of the analysis is to study the evolution of the
physical properties which dramatically affect galaxy spectra, such as the age,
metallicity, star formation history or dust content. More sophisticated
methods have been recently proposed \cite{HJL00,SSTL00}. Here I will
concentrate in describing the optimal parameter extraction method proposed by
\scite{HJL00}.

The main idea of the method in \scite{HJL00} is that, in practice, some of the
data may tell us very little about the parameters we are trying to estimate,
either through being very noisy, or through having no sensitivity to the
parameters. So in principle, we may be able to throw some data without loosing
much information about the parameters. It is obvious that throwing away data
is not the most optimal way. On the other hand, by performing linear
combinations of the data we will do better and then we can throw the linear
combinations which tell us least. Given a set of data {\bf x} (in our case
the spectrum of a galaxy) which includes a signal part ${\bf \mu}$ and noise
${\bf n}$, i.e. $\bx = \bbmu + \bn$, the idea then is to find a weighting
vector ${\bf b}$ such as $y \equiv {\bf b}^{t} {\bf x}$, it is these {\it
  numbers} $y$ which we are after.
 
In \scite{HJL00} an optimal and lossless method was found to calculate ${\bf
  b}$ for multiple parameters (as is the case with galaxy
  spectra). Specifically:

\begin{equation}
\bb_1 = {\bC^{-1} \bbmu_{,1}\over \sqrt{\bbmu_{,1}^t
\bC^{-1}\bbmu_{,1}}}
\label{Evector1}
\end{equation}

and

\begin{equation}
\bb_m = {\bC^{-1}\bbmu_{,m} - \sum_{q=1}^{m-1}(\bbmu_{,m}^t
\bb_q)\bb_q \over
\sqrt{\bbmu_{,m} \bC^{-1} \bbmu_{,m} - \sum_{q=1}^{m-1}
(\bbmu_{,m}^t \bb_q)^2}}.
\label{bbm}
\end{equation}

where a comma denotes the partial derivative with respect to the parameter
$m$ and $C$ is the covariance matrix with components $C_{ij}=<n_in_j>$.

The specific steps to build $m$ linear combinations to estimate $m$ parameters
are the following:

\begin{enumerate}

\item Choose a ``fiducial'' model (a first guess)

\item Compute the mean spectrum for the $m$ parameters and $m$ partial
  derivatives with respect to the $m$ parameters ($\bbmu_{,m}$).

\item Now compute $m$ eigenvectors $\bb_{i}$ from Eq.\ref{Evector1} and
  \ref{bbm}.

\item Finally, compute the $m$ $y_i$ values. This dataset is orthonormal, so
the new likelihood is easy to compute (the $y_m$ have mean
$\langle y_m \rangle = \bb_m^t \bbmu$ and unit variance), namely:

\be
\ln{\cal L}(\theta_\alpha) = {\rm constant} - \sum_{m=1}^{M}
{(y_m-\langle y_m\rangle)^2\over 2}
\ee

\end{enumerate}

\begin{figure}
\centerline{\psfig{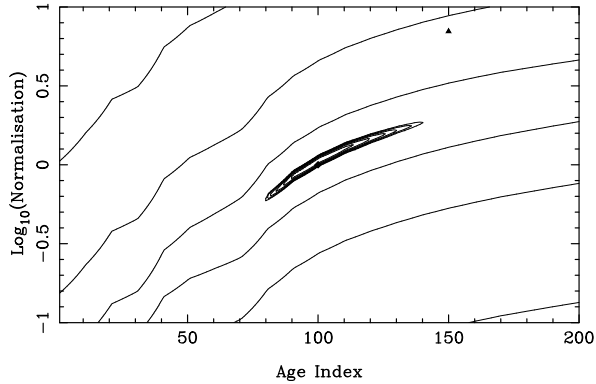}}
\caption{Full likelihood solution using all pixels.
  There are 6 contours running down from the peak value in steps of 0.5 (in
  $\ln{\cal L}$), and 3 outer contours at $-100$, $-1000$ and $-10000$.  The
  triangle in the upper-right corner marks the fiducial model which determines
  the eigenvectors to set the initial weights.}
\label{tth1}
\end{figure}
\begin{figure}
\centerline{\psfig{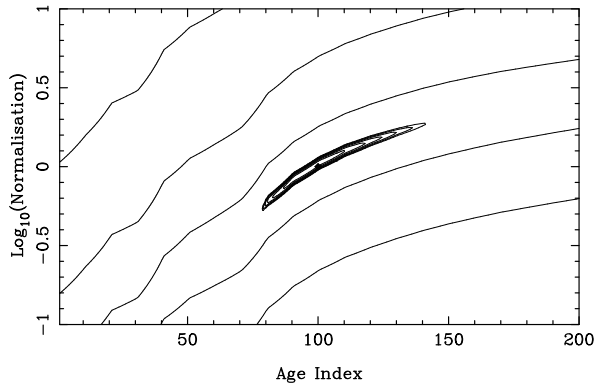}}
\caption{Likelihood solution using the compressed data set, i.e. the age 
datum and the normalization datum.  Contours are as in Fig. \ref{tth1}.}
\label{tth2}
\end{figure}

This procedure can be applied to derive the metallicities, ages, star
formation rates and dust content of galaxy spectra \cite{RJH00}.

It is very instructive to illustrate the method by trying to recover the age
and normalization of single stellar populations (SSP), i.e. the star formation
rate is $SFR(t') = A \delta(t'+t)$ where $\delta$ is a Dirac delta function.
The two parameters to determine are age $t$ and normalization $A$. We built a
simulated spectra with Gaussian noise and variance given by the mean, $\bC =
{\rm diag}(\bbmu_1, \ldots)$.  This is appropriate for photon number counts
when the number is large. It should be stressed that this is a more severe
test of the model than a typical galaxy spectrum, where the noise is likely to
be dominated by sources independent of the galaxy, such as CCD read-out noise
or sky background counts.  In the latter case, the compression method will do
even better than the example here (e.g. \scite{RJH00}). The simulated galaxy
spectrum is one of the galaxy spectra (age 3.95 Gyr, model number 100), and
the maximum signal-to-noise per bin is taken to be 2.  Noise is added,
approximately photon noise, with a Gaussian distribution with variance equal
to the number of photons in each channel.  Hence $\bC =$
diag($\bbmu_1,\bbmu_2,\ldots)$.  Figure~\ref{tth1} shows the contours in the
likelihood surface using all the points in the spectra. Figure~\ref{tth2}
shows the contours in the likelihood surface using {\it only two linear
  combinations: y$_1$ and y$_2$}. As it transpires from the figures, only two
numbers suffice to determine two parameters.

\section{The abundance of high--redshift objects}

Now that cosmic-microwave-background (CMB) experiments
\cite{boomerang3,maxima} have verified the inflationary
predictions of a flat Universe and structure formation from primordial
adiabatic perturbations, we are compelled to test further the predictions of
the simplest single-scalar-field slow-roll inflation models and to look for
possible deviations.  Measurements of the distribution of primordial density
perturbation afford such tests. The observed abundance of high--redshift
objects contains precious information about the properties of the initial
conditions. The reason for this is that the first objects to collapse, for a
given mass, will be due to fluctuations in the tail of the distribution of the
primordial density field and therefore will reflect the ``strength'' of it.
Furthermore, high--$z$ objects constrain the small scale part of the spectrum
of the primordial mass density field that cannot be probed directly by the
large scale structure of cosmic microwave background (CMB) observations.

\begin{figure}
\centerline{\psfig{figure=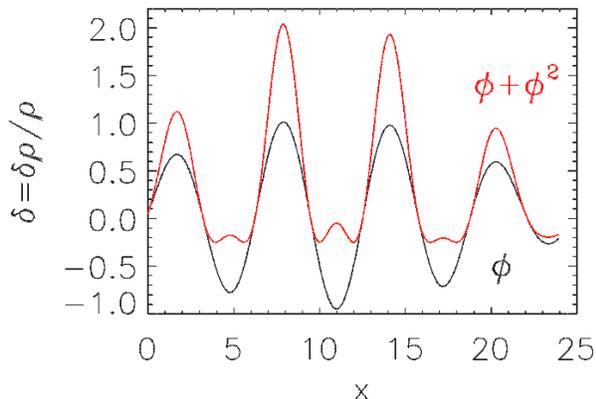,height=6cm,angle=0}}
\caption{The effect on the PDF for a  Gaussian field $\Phi$ of adding the
  square of itself. Note how the peaks get enhanced and the valleys suppressed.}
\label{gauss}
\end{figure}
 
The importance of using the mass-function as a tool to distinguish among
different non-Gaussian statistics for the primordial density field, was first
recognized by \scite{LM88,Colafrancescoetal89} and more recently, by
\scite{COS98}, followed by \scite{RB99}, \scite{RGS99a,RGS99b}, \scite{KST99},
\scite{Willick00}, \scite{AvelinoV99}. To make predictions on the number
counts of high-redshift structures in the context of non-Gaussian initial
conditions, a generalized version of the Press--Schechter (PS) theory has to
be introduced. The PS theory exploits the fact that in most cosmological
scenarios the large scale power exceeds that generated by non-linear coupling.
This in conjunction with specifying an ``artificial'' filtering of the initial
density field and a threshold for which we define objects that are able to
collapse, provides us with a description of the mass function in terms of the
probability density field (PDF) -- see \scite{Peacock99}. Thus in order to
extend the PS theory to the non-Gaussian case one needs to compute the
``smoothed'' PDF for the non-Gaussian field. Furthermore, numerical
simulations tell us that the PS theory provides a reasonable approximation for
the number of objects produced in tails -- provided we do not consider
fluctuations that deviate more than 5$\sigma$ from the mean where serious
deviations from the PS prediction occur
\cite{PS74,LeeShandarin98,ShethTormen99,Jenkinsetal00}.  Obtaining analytical
results in this context is extremely important. Direct simulations of
non-Gaussian fields are generally plagued by the difficulty of properly
accounting for the non-linear way in which resolution and finite box-size
effects, present in any realization of the underlying Gaussian process,
propagate into the statistical properties of the non-Gaussian field. Moreover,
finite volume realizations of non-Gaussian fields might fail in producing fair
samples of the assumed statistical distribution, i.e. ensemble and
(finite-volume) spatial distributions might sensibly differ.  This problem, of
course, becomes exacerbated and hard to keep under control in so far as the
tails of the distribution are concerned. Thus, in looking for the likelihood
of rare events for a non-Gaussian density field, either exact or approximate
analytical estimates should be considered as the primary tool.

\scite{RB99,RGS99a,RGS99b} considered a PDF which had a log-normal
distribution and assumed that it was the PDF which described the smoothed
field of fluctuations for a wide range of non--Gaussian models (mostly those
arising from structure formation by topological defects), based on comparison
s with numerical experiments. Their non-Gaussianity depends on a single
parameter, $G$ which is nothing but the ratio of 3$\sigma$ peaks in a
non-Gaussian model compared to the Gaussian case. An Einstein-deSitter
universe produces a noticeable deficit of high--redshift objects at
high--redshift (e.g. \scite{PJDWSSDW98}), RGS were able to find a region in
the $\sigma_8 - G$ plane for which the predicted cluster abundance in an EdS
universe agrees with observations (but see below).

\scite{Willick00} studied in great detail the mass determination of the
high--redshift cluster MS1054--03 concluding that its mass lies in the range
$1.4 \pm 0.3 \times 10^{15}$ M$_{\odot}$ for $\Omega_{\rm m}=0.3$. He then
investigated the amount of non-gaussianity needed to accommodate this cluster
within the CDM scenario using a parameterization for non-gaussianity similar
to that of \scite{RGS99a}. He found that MS1054--03 cannot be accommodated in
a CDM scenario with $\Omega_{\rm m} \ge 0.3$ unless some non-gaussianity
exists.

\scite{MVJ00} have computed an analytic expression for the probability
distribution function for a parameterization of primordial non-Gaussianity
that covered a wide range of physically motivated models: the non-Gaussian
field is given by a Gaussian field plus a term proportional to the square of a
Gaussian $\Phi=\phi+\epsilon_{\rm B}(\phi^2-\langle\phi^2\rangle)$, where
$\Phi$ applies to both the density perturbation field $\delta(\vec x)$ and the
primordial gravitational potential. They also introduced a generalized version
of the PS approach valid in the context of non--Gaussian initial conditions.
Note that \scite{MVJ00} considered only small departures from Gaussianity.
They also showed how this tiny departures can have a large impact in the
number density of observed objects at high--redshifts (see their Fig.~6).
Note also that if one considers large deviations from non-gaussianity, then
the normalization for $\sigma_8$ derived for Gaussian initial conditions is no
longer valid.

\begin{figure}
\centerline{\psfig{figure=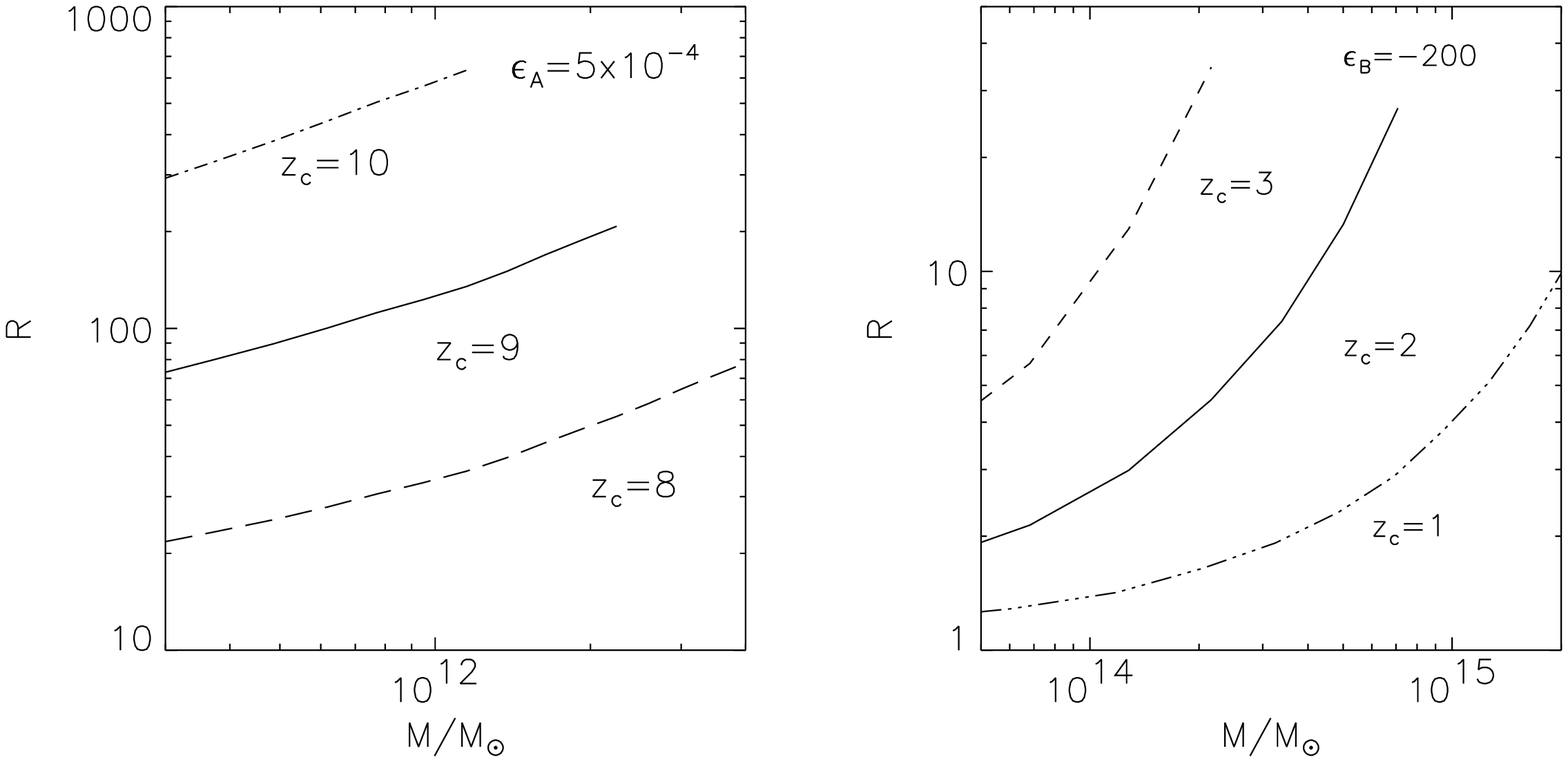,height=6cm,angle=0}}
\caption{Ratio $R(M,z)=N_{ng}(\geq M,z)/N(\geq M,z)$ for
  galaxies at redshift $z=8, 9$ and 10 for $\epsilon_{\rm A}=5\times 10^{-4}$
  (non-Gaussianity in the density field, left panel) and clusters at redshift
  $z=1 ,2$ and 3 (right panel), for $\epsilon_{\rm B}=200$ (non-Gaussianity in
  the potential) as a function of $M$. Lines are plotted only for masses
  where, for Gaussian initial conditions, one would expect to observe at least
  one object in the whole sky. Note that these high-redshift objects represent
  3- to 5-$\sigma$ peaks. The values for the number density enhancement $R$
  that can safely be attributed to primordial non-Gaussianity are $R=100$ for
  galaxies (left panel) and $R=10$ for clusters (right panel) }
\label{abund}
\end{figure}

\scite{VKMB00} have devised a method to constrain non-gaussianity by studying
the size-temperature distribution of galaxy clusters. The size-temperature
distribution is sensitive to the redshift of formation of the clusters. If
clusters originate from rare peaks of an initially Gaussian distribution, the
spread in formation redshift should be small and so should be the scatter in
the size-temperature distribution.  On the other hand, if the initial
distribution has long non-Gaussian tails, clusters we observe today should have
a broad formation redshift interval and therefore a large scatter in the
size-temperature distribution. They found that for the non-Gaussian parameters
derived by \scite{RGS99a} to explain the observed abundance of high-redshift
objects in an EdS universe, the spread in the size--temperature relation would
be much larger than is currently observed, thus excluding the possibility that
the EdS universe could be reconciled with observations of high--redshift
objects through a large amount of non-gaussianity. It is worth noting though
that this would not be the case for a non-gaussianity which comes from a
bimodal distribution with one of the modes centered at the cluster scale.

A comparison of the sensitivities for detecting non-gaussianity for several
tests has been investigated in \scite{VWHK00,VJKM00}. Using the kind of
non-gaussianity described in \scite{MVJ00}, they conclude that the CMB is
superior at finding non-Gaussianity in the primordial gravitational potential
(as inflation would produce), while observations of high--redshift galaxies
are much better suited to find non-Gaussianity that resembles that expected
from topological defects. Thus observations of high-redshift objects with the
Next Generation Space Telescope and the currently proposed 30--100 m class
telescopes should help us to shed light on the nature of the primordial
density field if -- and this is a big if -- mass determinations of these
objects can obtained with a 100\% error (see Fig.~\ref{abund}).

It is a pleasure to thank my collaborators in this work: Alan Heavens, Marc
Kamionkowski, Ofer Lahav, Sabino Matarrese and Licia Verde.


\begin{thebibliography}{{Reichardt}, {Jimenez} \& {Heavens}<2000>}
\bibitem[Avelino \& Viana<1999>]{AvelinoV99}Avelino P.~P., Viana P.,
  1999.\newblock {\rm preprint}, {\rm astro-ph}, 9907209.
\bibitem[Bromley {\rm et~al.}<1998>]{Bromley98}Bromley B., Press W., Lin H.,
  Kirschner R., 1998.\newblock {\rm ApJ}, {\rm 505}, 25.
\bibitem[Chiu, Ostriker \& Strauss<1998>]{COS98}Chiu W.~A., Ostriker J.,
  Strauss M.~A., 1998.\newblock {\rm ApJ}, {\rm 494}, 479.
\bibitem[Colafrancesco, Lucchin \&
  Matarrese<1989>]{Colafrancescoetal89}Colafrancesco S., Lucchin F., Matarrese
  S., 1989.\newblock {\rm ApJ}, {\rm 345}, 3.
\bibitem[Connolly \& Szalay<1999>]{Connolly99}Connolly A., Szalay A.,
  1999.\newblock {\rm AJ}, {\rm 117}, 2052.
\bibitem[Connolly {\rm et~al.}<1995>]{Connolly95}Connolly A., Szalay A.,
  Bershady M., Kinney A., Calzetti D., 1995.\newblock {\rm AJ}, {\rm 110},
  1071.
\bibitem[{de Bernardis et al.}<2000>]{boomerang3}{de Bernardis et al.},
  2000.\newblock {\rm nature}, {\rm 404}, 955.
\bibitem[Folkes {\rm et~al.}<1999>]{Folkes99}Folkes et~al., 1999.\newblock {\rm MNRAS}, {\rm 308}, 459.
\bibitem[Folkes, Lahav \& Maddox<1996>]{Folkes96}Folkes S., Lahav O., Maddox
  S., 1996.\newblock {\rm MNRAS}, {\rm 283}, 651.
\bibitem[Francis {\rm et~al.}<1992>]{Francis92}Francis P., Hewett P., Foltz C.,
  Chaffee F., 1992.\newblock {\rm ApJ}, {\rm 398}, 476.
\bibitem[Galaz \& deLapparent<1998>]{Galaz98}Galaz G., de~Lapparent V.,
  1998.\newblock {\rm A\& A}, {\rm 332}, 459.
\bibitem[Glazebrook, Offer \& Deeley<1998>]{Glazebrook98}Glazebrook K., Offer
  A., Deeley K., 1998.\newblock {\rm ApJ}, {\rm 492}, 98.
\bibitem[{Heavens}, {Jimenez} \& {Lahav}<2000>]{HJL00}{Heavens} A., {Jimenez}
  R., {Lahav} O., 2000.\newblock {\rm MNRAS}, {\rm 317}, 965.
\bibitem[{Jaffe et~al.}<2000>]{maxima}{Jaffe et~al.}, 2000.\newblock {\rm Phys.
  Rev. Lett}, {\rm astro-ph}, 0007333.
\bibitem[Jenkins {\rm et~al.}<2000>]{Jenkinsetal00}Jenkins A., Frenk C.~S.,
  White S. D.~M., Colberg J.~M., Cole S., Evrard A.~E., Yoshida N.,
  2000.\newblock {\rm MNRAS}, {\rm astro-ph}, 0005260.
\bibitem[Koyama, Soda \& Taruya<1999>]{KST99}Koyama K., Soda J., Taruya A.,
  1999.\newblock {\rm astro-ph}, {\rm 9903027}.
\bibitem[Lee \& Shandarin<1998>]{LeeShandarin98}Lee J., Shandarin S.~F.,
  1998.\newblock {\rm ApJ}, {\rm 500}, 14.
\bibitem[Lucchin \& Matarrese<1988>]{LM88}Lucchin F., Matarrese S.,
  1988.\newblock {\rm ApJ}, {\rm 330}, 535.
\bibitem[Matarrese, Verde \& Jimenez<2000>]{MVJ00}Matarrese S., Verde L.,
  Jimenez R., 2000.\newblock {\rm ApJ}, {\rm 541}, 10.
\bibitem[Murtagh \& Heck<1987>]{Murtagh87}Murtagh F., Heck A., 1987.\newblock
  {\it Multivariate Data Analysis}, Reidel, Dordrecht.
\bibitem[{Peacock} {\rm et~al.}<1998>]{PJDWSSDW98}{Peacock} J.~A., {Jimenez}
  R., {Dunlop} J.~S., {Waddington} I., {Spinrad} H., {Stern} D., {Dey} A.,
  {Windhorst} R.~A., 1998.\newblock {\rm MNRAS}, {\rm 296}, 1089.
\bibitem[Peacock<1999>]{Peacock99}Peacock J.~A., 1999.\newblock {\it
  Cosmological Physics}, Cambridge University Press.
\bibitem[Press \& Schechter<1974>]{PS74}Press W.~H., Schechter P.,
  1974.\newblock {\rm ApJ}, {\rm 187}, 425.
\bibitem[{Reichardt}, {Jimenez} \& {Heavens}<2000>]{RJH00}{Reichardt} C.,
  {Jimenez} R., {Heavens} A.~F., 2000.\newblock {\rm astro-ph}, {\rm }.
\bibitem[Robinson \& Baker<1999>]{RB99}Robinson J., Baker J., 1999.\newblock
  {\rm ApJ}, {\rm astro-ph}, 9905098.
\bibitem[Robinson, Gawiser \& Silk<1999a>]{RGS99a}Robinson J., Gawiser E., Silk
  J., 1999a.\newblock {\rm ApJ(Lett)}, {\rm astro-ph}, 9805181.
\bibitem[Robinson, Gawiser \& Silk<1999b>]{RGS99b}Robinson J., Gawiser E., Silk
  J., 1999b.\newblock {\rm ApJ}, {\rm astro-ph}, 9906156.
\bibitem[Ronen, Aragon-Salamanca \& Lahav<1999>]{Ronen99}Ronen R.~T.,
  Aragon-Salamanca A., Lahav O., 1999.\newblock {\rm MNRAS}, {\rm 303}, 284.
\bibitem[Sheth \& Tormen<1999>]{ShethTormen99}Sheth R., Tormen G.,
  1999.\newblock {\rm MNRAS}, {\rm 308}, 119.
\bibitem[Singh, Gulati \& Gupta<1998>]{Singh98}Singh H., Gulati R., Gupta R.,
  1998.\newblock {\rm MNRAS}, {\rm 295}, 312.
\bibitem[{Slonim} {\rm et~al.}<2000>]{SSTL00}{Slonim} N., {Somerville} R.,
  {Tishby} N., {Lahav} O., 2000.\newblock {\rm astro-ph}, {\rm 0005306}.
\bibitem[{Verde} {\rm et~al.}<2000a>]{VJKM00}{Verde} L., {Jimenez} R.,
  {Kamionkowski} M., {Matarrese} S., 2000a.\newblock {\rm astro-ph}, {\rm }.
\bibitem[{Verde} {\rm et~al.}<2000b>]{VKMB00}{Verde} L., {Kamionkowski} M.,
  {Mohr} J., {Benson} A., 2000b.\newblock {\rm astro-ph}, {\rm 0007426}.
\bibitem[{Verde} {\rm et~al.}<2000c>]{VWHK00}{Verde} L., {Wang} L., {Heavens}
  A.~F., {Kamionkowski} M., 2000.\newblock {\rm MNRAS}, {\rm 313}, 141.
\bibitem[Willick<2000>]{Willick00}Willick J.~A., 2000.\newblock {\rm ApJ}, {\rm
  530}, 80.

\end{thebibliography}

\end{document}